\documentstyle[A4,12pt,epsf]{article}
\begin{document}
\begin{titlepage}
\vspace{2cm} 

\begin{flushright}
PSI-PR-96-07\\
hep-ph/9602340
\end{flushright}

\vspace{1cm}
\begin{center}
 
\begin{large}
{\bf A Renormalization Group Analysis of the Higgs Boson with  
Heavy Fermions and Compositeness}
\end{large}

\vspace*{2cm}
	{\bf Hanqing Zheng$^\dagger$ }\\
\vspace*{1.0cm}
	Paul Scherrer Institute, 5232 Villigen PSI, Switzerland

\end{center}
\vspace*{2cm}
\begin{abstract}

We study the properties of heavy fermions in the 
vector-like representation of the electro-weak gauge group 
$SU(2)_W\times U(1)_Y$ with Yukawa couplings to the standard model 
 Higgs boson. 
Using the renormalization group analysis, we 
discuss  their effects on 
the vacuum stability
and the triviality bound on the Higgs self-coupling, 
within the context of the standard model (i.e., the Higgs
particle is elementary).  Contrary to the low energy case
where the decoupling theorem dictates their behavior, the  inclusion of 
heavy fermions drastically changes the standard model
 structure at high scales.
We also discuss the interesting possibility of compositeness,
i.e., the Higgs particle is  composed of the heavy fermions using the method
of Bardeen, Hill and Lindner~\cite{BHL91}.
Finally we briefly comment on their possible role in 
explaining $R_b$ and $R_c$.
\end{abstract}

\vfill
\vfill

\begin{flushleft}
$\dagger$) e-mail address: hanqing.zheng@psi.ch
\end{flushleft}

\phantom{p}
\vfill
\phantom{p}
\end{titlepage}
Enormous efforts have been made in
searching for physics beyond the standard model but up to now a crucial,
direct
experimental indication is still illusive. One of the most important
motivation to explore heavy fermions above the energy accessible by
current accelerators is to look for extra building blocks of nature
beyond the three families of the standard model.  
For this purpose it may be adequate to look at fermions in vector-like
representations of the electro-weak 
gauge group with a large bare mass term,
rather than the conventional chiral doublets.
The main reason for this is from the strong experimental constraint
on the S parameter~\cite{peskin}. While the experiments favour a 
negative value of S~\cite{exp},
a standard chiral doublet of heavy fermions (degenerate in mass) contributes
to the S parameter as $1/6\pi$.
On the contrary, for fermions in the
vector-like representation of the electro-weak gauge group a large bare 
fermion mass  M completely changes the low energy properties of the 
heavy fermions. As a consequence of the decoupling theorem, heavy fermions
contribute to the S and T parameters as\footnote{This is also
true when M
is generated by chiral dynamics at higher scale but irrelevant to 
electro-weak physics. For the definition of the T parameter and other low
energy observables, see also ref.~\cite{peskin}.},
\begin{equation}
S\sim {m^2\over M^2}, \,\,\,\,\, 
T\sim {m'^2\over M^2}\cdot{m'^2\over v^2},
\end{equation}
where $m$ ($m'$) is the  mass parameter generated by the 
Yukawa coupling responsible for the electro-weak symmetry breaking 
(weak isospin violation), $v$(=246~GeV) is the vacuum
expectation value of the Higgs particle. 
The interesting aspect
of  the vector-like heavy fermions is not that they are safe
at low energies, but most importantly, because they may have
some interesting
role in physics beyond the standard model. For example, they may be
responsible for a dynamical generation of 
light fermion mass matrix~\cite{nielsen}, 
they are natural consequences of 
many
grand unification models,
and they 
may even be the constituents of the electro-weak Higgs 
particle~\cite{vhiggs}. 
Therefore it is important to investigate the fundamental
properties of the heavy vector-like fermions and this is the purpose of the 
current paper. 
The model we will study in this paper 
is the minimal standard model plus these hypothetical heavy vector-like
fermions with Yukawa interactions to the Higgs boson. 
In most cases bosons have
just the opposite property to that of fermions. 
To be specific, in this paper we assume
no elementary boson exists except for the known gauge bosons (and maybe
the Higgs particle), in particular we assume no super-symmetry.   

Recently there has been a
 renewed interest in understanding the structure of the
standard model at high energies, even up to Planck 
scale~\cite{stab,CEQ,vac}. 
A powerful
tool is to use the renormalization group equations to trace the evolution 
of the coupling constant of the $\lambda \phi^4$
self-interaction of the Higgs particle. 
Assuming the standard model remains
valid up to certain scale $\Lambda$, 
an upper bound (the triviality bound,
obtained by requiring
 $\lambda$ not to blow up below $\Lambda$) of the Higgs boson mass,
$m_H$, can be obtained. Meanwhile, 
requiring the stability
of the  electro-weak vacuum, we can also obtain a lower bound
on $m_h$. For the later purpose, in principle one needs to 
consider the
renormalization group improved effective potential~\cite{sher} 
and require it bounded from below. But in practice
this turns out to be equivalent to the 
requirement that the Higgs self-interaction
coupling constant $\lambda$ does not become negative, below the given
scale (see \cite{stab} and ref. therein). 
It is remarkable to note that for the given experimental value of the top 
quark mass (here we use $m_t=174$GeV), 
there is an allowed range for the Higgs boson mass,
$130\hbox {GeV}\leq$$m_H$$\leq 200\hbox{GeV}$, 
for which the standard model may remain valid up to Planck
scale.

For the purpose of having  a qualitative understanding
of the influence of heavy fermions to the above analysis, 
it is enough to 
consider the  following simple Lagrangian,
\begin{equation}
\label{L}
{\cal L}=\bar Q(i {/\llap D}_d-M)Q + 
\bar U(i {/\llap D}_s-M)U +\bar D(i {/\llap D}_s-M)D + 
\{g_Y\bar Q\phi D + g_Y'\bar Q\tilde\phi U + h.c.\}   \ .
\end{equation} 
In  above Lagrangian we introduced four vector-like fermions,
$Q$ is a $SU(2)_W$  doublet  and $U$ and $D$ are singlets. 
We assume
they participate in strong interactions
and are in fundamental representations of $SU(3)_C$. 
They are equivalent to one family 
of chiral quarks plus a left--right conjugated  chiral quark family.
We call them one family of vector-like quarks.
The subscript  $d$ ($s$) in the covariant 
derivatives denotes that the corresponding fermion is a
$\hbox{SU(2)}_W$ doublet (singlet)
and $\phi$ denotes the standard Higgs doublet. We further expect the Yukawa
couplings ($g_Y$ and $g_Y'$) to be of order 1.
For simplicity we take all the 
bare fermion masses to be equal. 

When heavy 
fermions are included the structure
of our world changes drastically at high scales, 
even though vector-like
fermions are essentially decoupling below their thresholds. At scales
much higher than the threshold whether the fermion field is chiral or
vector-like does not make any qualitative difference. The 
picture obtained from the following analysis also holds for 
Yukawa couplings between chiral quarks or between 
chiral and vector-like quarks.

	The correction to the effective potential 
from the heavy fermion is
$$\delta V(\phi_c)= -{N_c\over 16\pi^2}\{
(M+g_Y\phi_c)^4\log({(M+g_Y\phi_c)^2\over\mu^2})
+(M-g_Y\phi_c)^4\log({(M-g_Y\phi_c)^2\over\mu^2})\}
$$
\begin{equation}  
+(g_Y\to g_Y') \ .
\end{equation}
As is well known,
because of the negative sign,
fermions turn to destabilize the vacuum.
At a scale $\phi_c<M$ one can expand the above expression in powers of
$\phi_c^2/ M^2$ and 
it is easy to verify that heavy fermions decouple from the
effective potential as a consequence of the decoupling theorem. Far 
above the threshold there is no difference between chiral and vector-like
fermions. The only  essential ingredient 
is the number of independent Yukawa 
couplings (notice that $\phi_c\bar Q U$ 
and $\phi_c\bar U Q$ are counted as
different).\footnote{There is another theoretical approach in the literature
which can be related to the studies on  Yukawa couplings: 
the Veltman condition. However in the Veltman condition there are no 
decoupling or  threshold effects. The condition simply counts the number 
of independent Yukawa couplings (and of course, their strength). 
In this paper, we do not investigate this approach.}

In the following we list
the relevant renormalization group equations 
(for simplicity we neglect the
$SU(2)_W\times U(1)_Y$ couplings $g_2$ and $g_1$, since 
their effects are relatively
small and do not change the qualitative picture):
\begin{equation}
\label{l}
16\pi^2{d\lambda\over dt}= 24\lambda^2+12\lambda (g_t^2+4N_dg_Y^2)
-6(g_t^4+4N_dg_Y^4) \ ,
\end{equation}
\begin{equation}
16\pi^2{dg_Y\over dt}= {(24N_d+3)\over 2}g_Y^3-8g_s^2g_Y\ ,
\end{equation}
\begin{equation}
16\pi^2{dg_s\over dt}= -(21-8N_c\theta)/3g_s^3
\end{equation}
where $g_t$ is the Yukawa coupling of the top quark 
($g_t=\sqrt{2}m_t/v$) and $g_s$ is the strong coupling constant.
The beta function for top quark Yukawa coupling is the same as the standard
model one at 1 loop we therefore do not list it here.
We take $g_Y=g_Y'$ for simplicity.
The renormalization group equations are
written in a more general case:
$4N_d$ is 
the number of independent Yukawa couplings.
$4N_c$ is the number of colored heavy fermions.
For example $N_d=1/2$ 
corresponds to setting $g_Y$ (or $g_Y'$) in eq.~(\ref{L})
to zero.
We use a simple  step
function $\theta=\theta(t-log(M/M_z))$ to model the 
heavy fermion threshold effects. 
All the $g_Y$ couplings in above renormalization group 
equations are understood
as multiplied by $\theta$.

In fig.~\ref{fig1} we plot the vacuum
stability bound and the triviality bound on the Higgs mass as a 
function of the 
scale $\Lambda$ for some typical values of the 
parameters of the  heavy fermions.
We see that the inclusion of heavy fermions drastically change the 
Standard model structure at high energies 
even though they decouple from 
the low energy world.
They always tighten the bound on the 
mass of the Higgs boson as a function of the cutoff scale $\Lambda$.
Notice that in principle 
(in terms of one loop renormalization equations)
the upper line (triviality bound) and
the lower line (vacuum stability bound) never meet each other. 
Because the
upper line is drawn by requiring $\lambda$ not to blow up and the 
lower line is drawn by requiring $\lambda\geq 0$. 
Between them is the ultra-violet unstable fixed point of $\lambda$, 
so the two lines get 
close to each other rapidly.

We now study the interesting possibility of considering 
the Higgs particle as
a composite object of the heavy vector-like fermions. 
Assuming the dynamical symmetry 
breaking occurs through a mechanism 
{\it \`a la} 
Nambu--Jona-Lasinio (via a four--fermi interaction) or through an
effective Higgs--Yukawa interaction,
it is proven in
\cite{vhiggs} that, after 
integrating out the heavy fermion degrees of freedom,
the model is completely equivalent to the standard model at low energies 
and the Higgs boson's mass has nothing to do with the bare fermion mass,
even though the
Higgs boson is composed of heavy fermions. 
Applying the above renormalization group analysis
to the composite model leads  
to some interesting results which we 
present below. We follow the method 
proposed by Bardeen, Hill and Lindner
(BHL)
\cite{BHL91} originally developed for the top quark condensate model.
The  basic idea of the BHL method is 
the following: Using the collective field method
 the four--fermi interaction Lagrangian
is rewritten into an effective Higgs--Yukawa
interaction Lagrangian
at the cutoff scale $\Lambda$ (the coupling strength of the
four-fermi interaction is $\sim G/\Lambda^2$,
where $G$ is a dimensionless
coupling constant). The effective Yukawa interaction 
Lagrangian is identical to the standard model at the
cutoff scale $\Lambda$, but
with vanishing wave function renormalization constant of the Higgs
field ($Z_H=0$) and vanishing Higgs self-coupling ($\lambda=0$). 
Below $\Lambda$ the model is equivalent to the standard model and
therefore the coupling constants of the effective 
theory run according to the
standard model renormalization group equations. However the
vanishing
 of $Z_H$ at the scale $\mu=\Lambda$ 
leads to the following boundary
conditions of the renormalization group equations:
\begin{equation}
g_Y^r \to \infty \ , \,\,\,\, \lambda^r/(g^r_Y)^4 \to 0 
\end{equation}
where $\lambda^r$ and $g^r_Y$ are the renormalized Higgs self-coupling and
Yukawa coupling, respectively. With the renormalization group equations
 and
boundary conditions, one can predict the 
mass of the Higgs boson and the fermion mass 
(or the Yukawa coupling) at its infra-red fixed point value.
In the present paper, of course, 
the ``standard model" often refers to the standard model plus
heavy fermions and the ``infra-red fixed point" value of $g_Y$ refers to 
its value at the heavy fermion threshold.

The minimal top quark 
condensate model has already been ruled out by
experiments. In order to generate the electroweak 
symmetry breaking scale $v$,
the top quark mass is required to be at least as large as 218~GeV 
(corresponding to $\Lambda=10^{19}$~GeV, i.e., Planck scale).  
The experimental bound on the top quark mass ($180\pm 13$~GeV
\cite{top}) indicates that the top
quark Yukawa coupling does not diverge  up to Planck scale
in the standard model  and therefore does
not meet the compositeness condition of BHL (see however \cite{hasen}).
Since in the present model there is no strict experimental constraint,
the 
compositeness condition is easily and naturally achievable. 
In fig.~\ref{fig2}
we plot the 
composite Higgs particle's 
mass\footnote{The Higgs mass in these figures is the
renormalized mass at $\mu=M_Z$. The renormalized mass
is close to the pole mass of the Higgs boson.} and 
the Yukawa coupling $g_Y$ at 
the infra-red fixed point, as a function of $\Lambda$.
For a large number of colored fermions
there is a problem with non-asymptotic freedom of the strong gauge coupling.
Non-asymptotic freedom
already happens with 3 vector-like quark families. However for
$N_c=3$ the strong coupling constant behaves mildly up to Planck scale and
causes no problem.
For simplicity we set all the Yukawa couplings 
to be equal. 

From fig.~\ref{fig2}
we see that the allowed range
for the Higgs mass is rather narrow 
against the wide range of the cutoff scale,
the bare fermion mass  and the number of families except when the 
heavy fermion bare mass $M$     is 
 close to the cutoff $\Lambda$. A  lower bound on the Higgs mass%
can be obtained:
 $m_H\geq 145$~GeV. When $M$ is getting close to the cutoff
scale our results become  unstable and are sensitive to
the input numerical values of the boundary conditions. In such a
situation the scale is not large enough for the couplings to reach the
infra-red stable point. However
it is  estimated that the Higgs mass
will not exceed 450~GeV, 
otherwise the whole mechanism become unnatural (in the
sense that the Yukawa coupling constant at electroweak scale also becomes
substantially larger than 1).

In fig.~\ref{fig3}
we plot a typical example of the Higgs mass for a given
 cutoff scale $\Lambda_c$ and $N_d$ ($N_c$). 
We also plot the triviality bound and the
vacuum stability bound using the Yukawa coupling constant at its infrared
fixed-point value as the initial boundary condition,
 which is determined
uniquely by the parameter $M$,
$\Lambda_c$ and $N_d$ ($N_c$) in the compositeness picture. 
It is very interesting to notice 
that $m_H$ and $\Lambda$ take the values where
the curves of triviality bound and vacuum stability bound (practically) meet
each other. This is 
the unique feature of BHL compositeness picture.
The reason behind this is very simple: The 
infra-red attractive 
fixed point corresponds to
the ultra-violet unstable fixed point. 
In the sense  of ref.~\cite{hasen}, this
picture can be disturbed. However in most cases the infra-red--ultra-violet
fixed point structure is influential and 
rather stable against perturbation.
From the above analysis,
it is clear that the Higgs particle's mass obtained in a 
composite model will not contradict~\footnote{That is, the low energy model
should not blow up before reaching the compositeness scale.}
the triviality bound obtained
from  its  low energy theory. 
Without a renormalization group analysis, 
the obtained Higgs mass in the limit of large number of colors 
like the relation $m_H=2m_t$ 
in top condensation model or that $m_H$ equals to the mass splitting 
between heavy vector-like fermions in the current model does often 
contradict the triviality bound\footnote{For the discussion
on the relations between the 
Higgs boson mass obtained from the renormalization group analysis and 
that from a large $N_c$ analysis, see \cite{BDL95}.}.

Recently several authors~\cite{cck} have suggested that a proper
 mixing between heavy vector-like fermions 
and the $b,c$ quarks may resolve the $R_b$
($\equiv \Gamma_b/\Gamma_{hadron}$) and 
$R_c$ ($\equiv \Gamma_c/\Gamma_{hadron}$) crisis. 
In the second model of  Chang, Chang and Keung's paper in ref.~\cite{cck},
one weak singlet, charge 2/3 heavy quark 
which mixes 
with the charm quark is introduced to reduce $R_c$, 
and a vectorial $SU(2)_W$ doublet with weak hypercharge 
$- 5/3 $
mixing with the down--type quarks is introduced
to increase $R_b$. The influence on $R_c$ and $R_b$ depends 
on the mixing angle between the vectorial and the chiral quarks. 
The mixing angle, $\theta\sim  g_Yv/\sqrt{2}M$ can be
determined from the experimental value of $R_b$ and $R_c$. 
For sufficiently large $M$, for example, $M\sim 1$~TeV
\footnote{$M$ cannot be taken
too large in order to be consistent with the experimental constraints on the 
T parameter. Fixed mixing angle requires the increasing of the 
Yukawa coupling when $M$ is increased.}, 
a rough estimate indicates that
the Yukawa coupling is already large enough to meet the 
compositeness condition (i.e., the running Yukawa coupling blows up below 
the Planck scale.). However a more quantitative analysis requires considering 
also the running of $g_2$ and $g_1$ couplings.  

 {\it Acknowledgement}: I thank H. Schlereth for his interest  
and discussions on the topic of this paper. I also would like to 
thank Frank Cuypers for a careful reading of the manuscript and M. Lindner
for several interesting comments.

\begin{newpage}
\begin{figure}[hbtp]
\begin{center}
\mbox{\epsfysize=120mm\epsffile{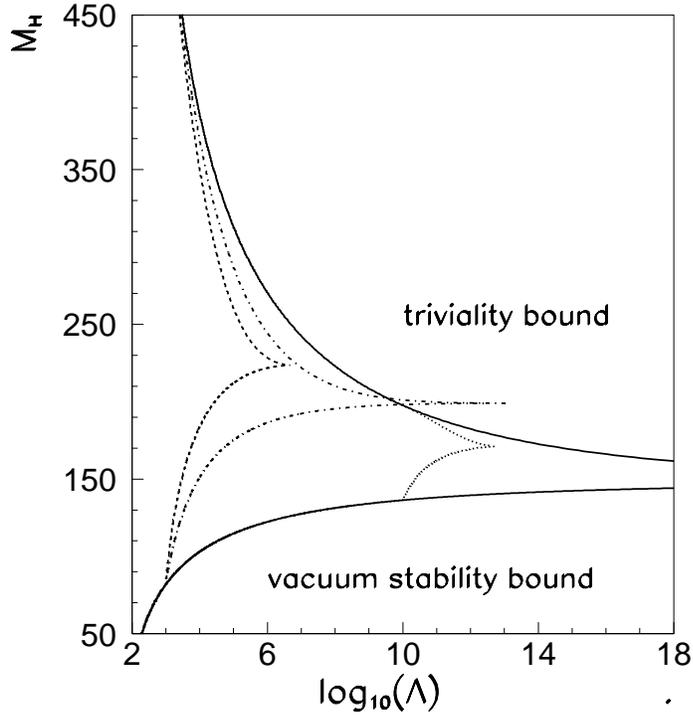}}
\vspace*{-5mm} 
\caption{\label{fig1}%
Vacuum stability and  triviality bounds on the Higgs mass 
as a function of $\Lambda$.
The solid lines are the standard model case 
(in the absence of $g_2$ and $g_1$, we take $\alpha_s=0.118$); 
The dot-dashed lines correspond to 
$N_d=1/2$, $N_c=3/4$ (i.e., 
without ``$U$'' or ``$D$'' quark in eq.~(\protect\ref{L})) and  
$M=10^3$~GeV; The dashed (dotted) lines correspond 
to $N_d=1$, $N_c=1$ and 
$M=10^3$~GeV ($M=10^{10}$~GeV), respectively. The Yukawa coupling 
$g_Y=1$.}
\end{center}
\end{figure}

\newpage
\begin{figure}[hbtp]
\begin{center}
\vspace*{-25mm} 
\mbox{\epsfysize=70mm\epsffile{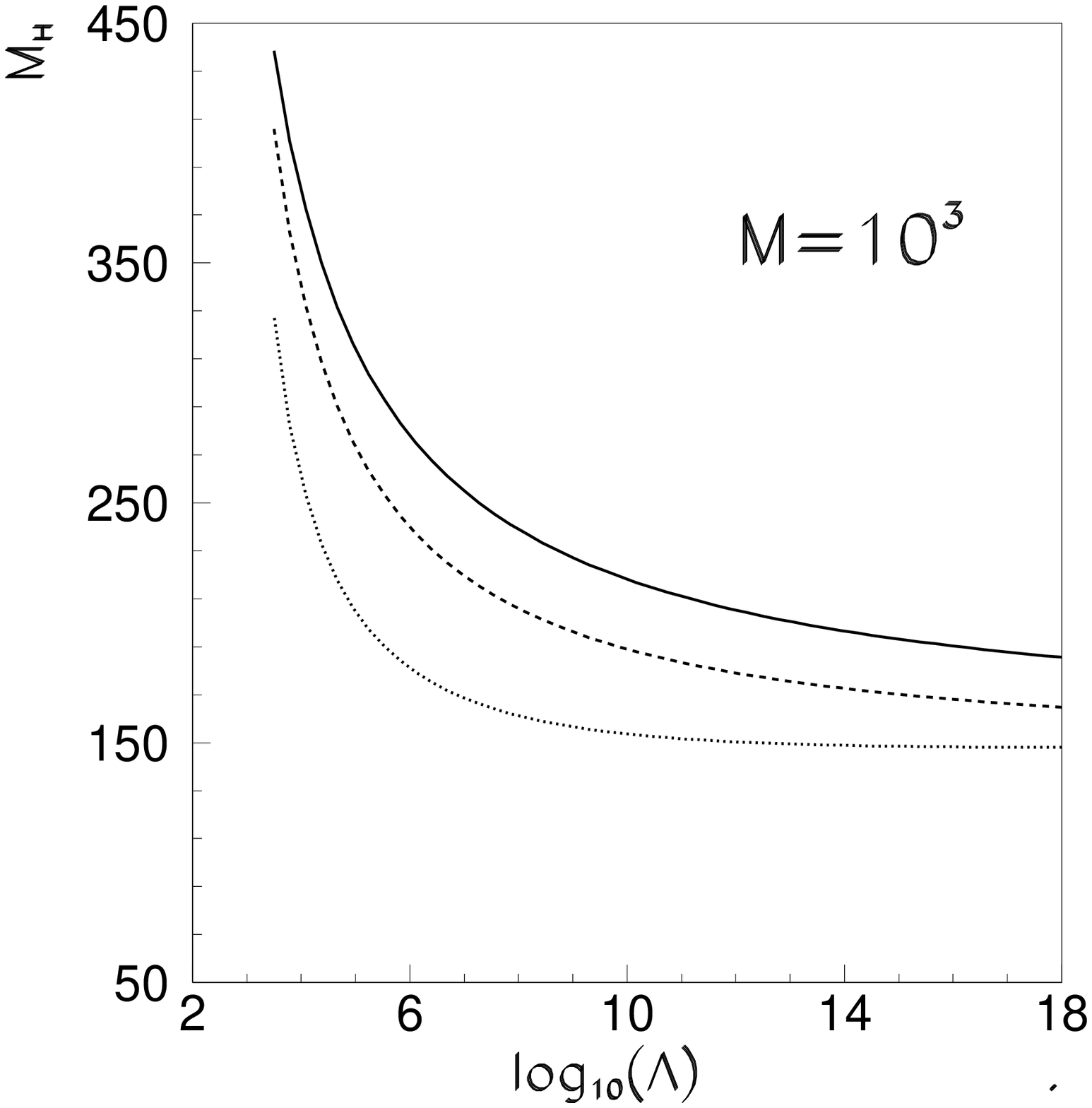}}\hspace*{10mm}
\mbox{\epsfysize=70mm\epsffile{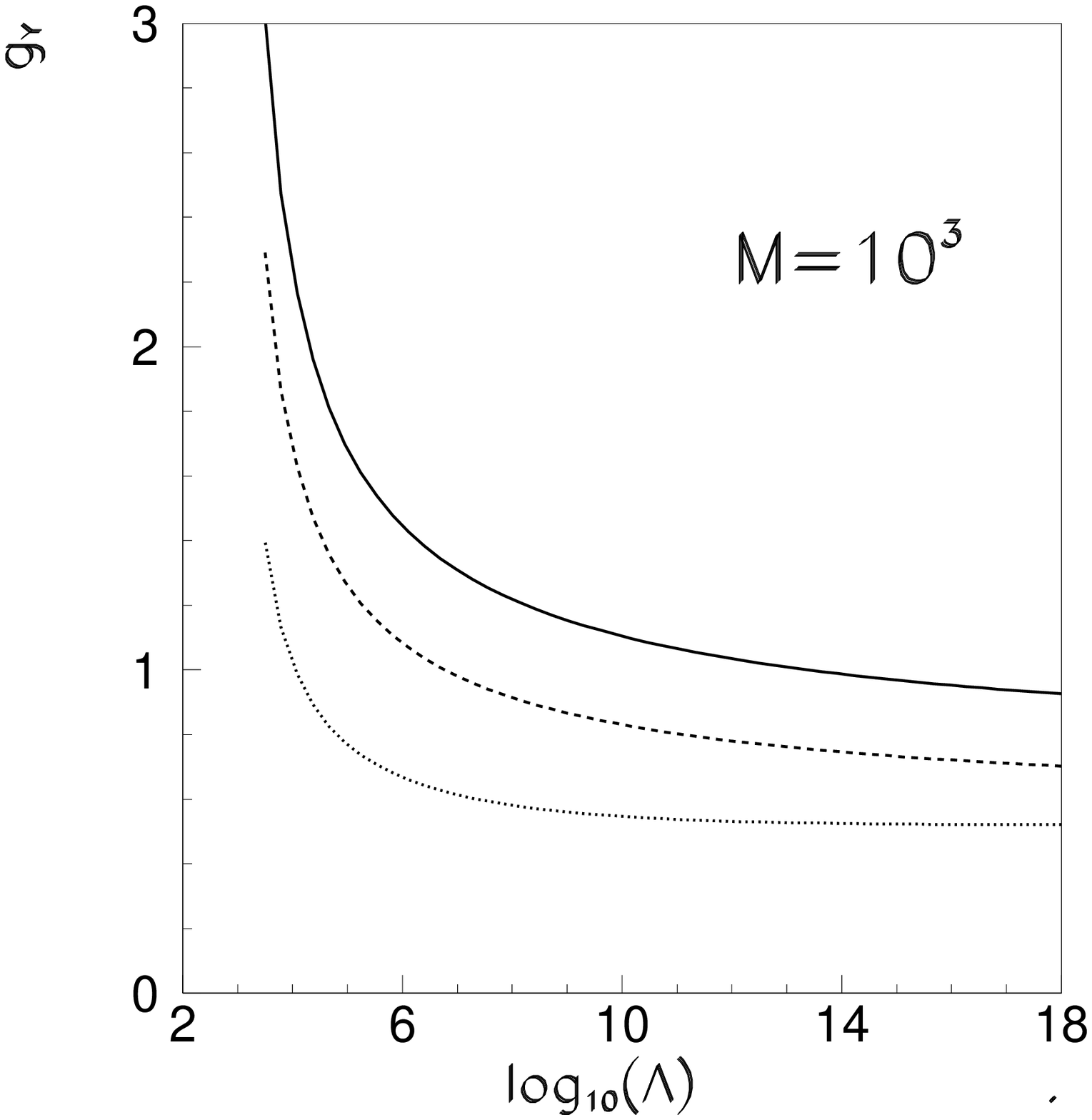}}
\vspace*{-5mm} 
\mbox{\epsfysize=70mm\epsffile{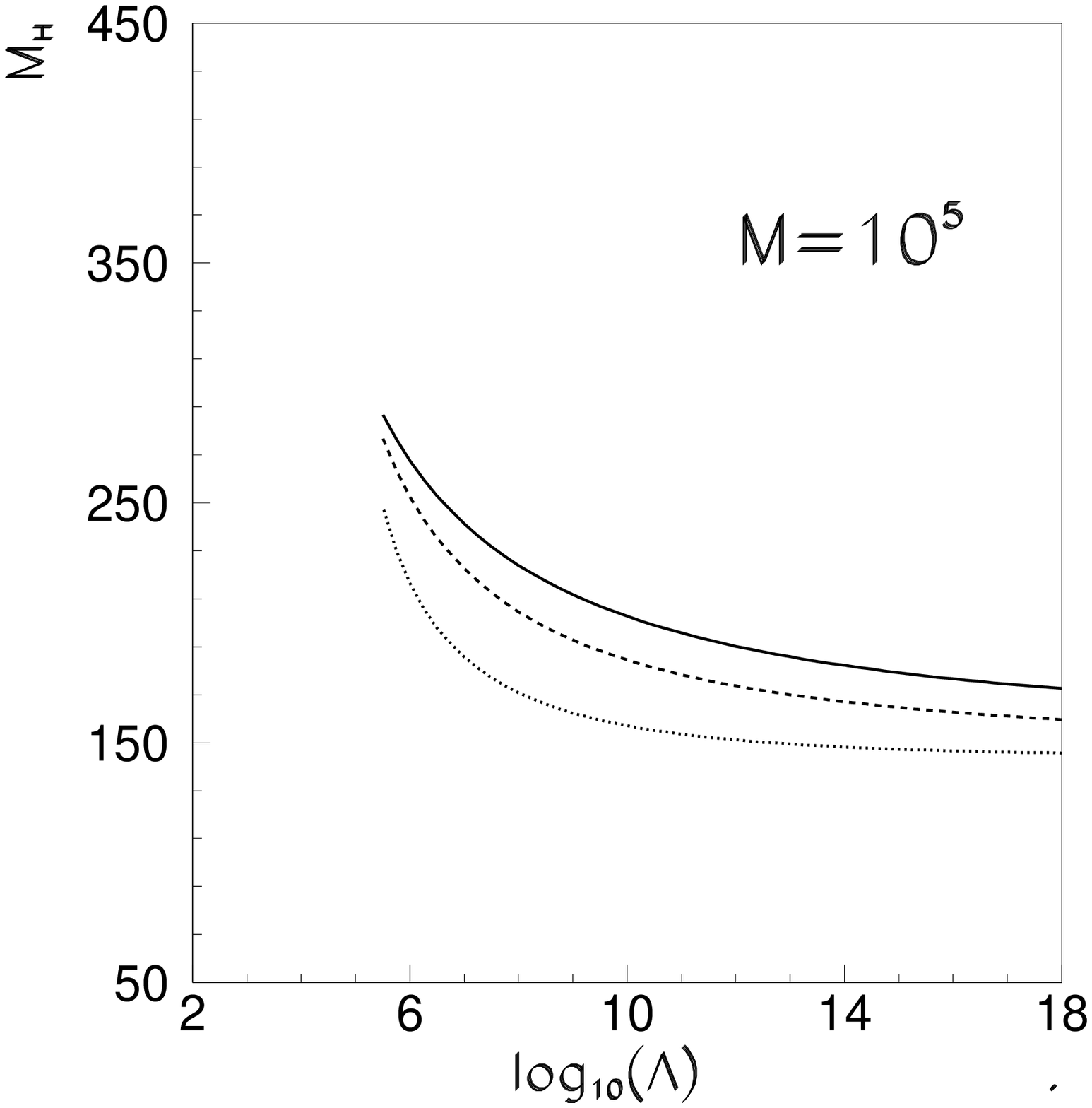}}\hspace*{10mm}
\mbox{\epsfysize=70mm\epsffile{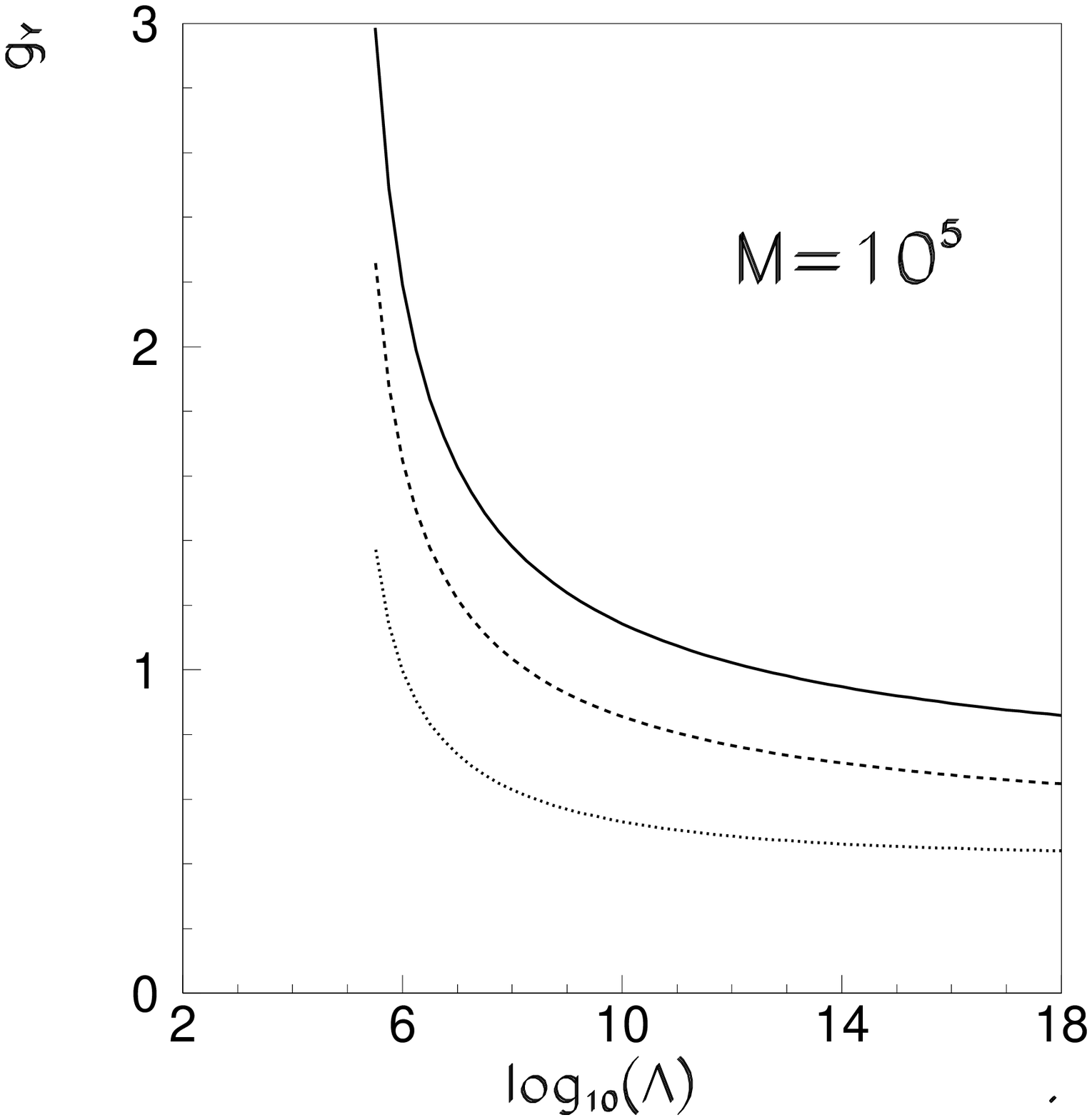}}
\caption{ \label{fig2}%
IR fixed point
value (at $M_Z$) of $M_H$ and $g_Y$ as functions of 
the compositeness scale. 
The solid line: $N_d=1/2$, $N_c=3/4$;
the dashed  line: $N_d=N_c=1$; the dotted line: $N_d=N_c=3$.
The value of $M$ in the figure is in unit of GeV.}
\end{center}
\end{figure}

\newpage
\begin{figure}[hbtp]
\begin{center}
\vspace*{-50mm} 
\mbox{\epsfysize=120mm\epsffile{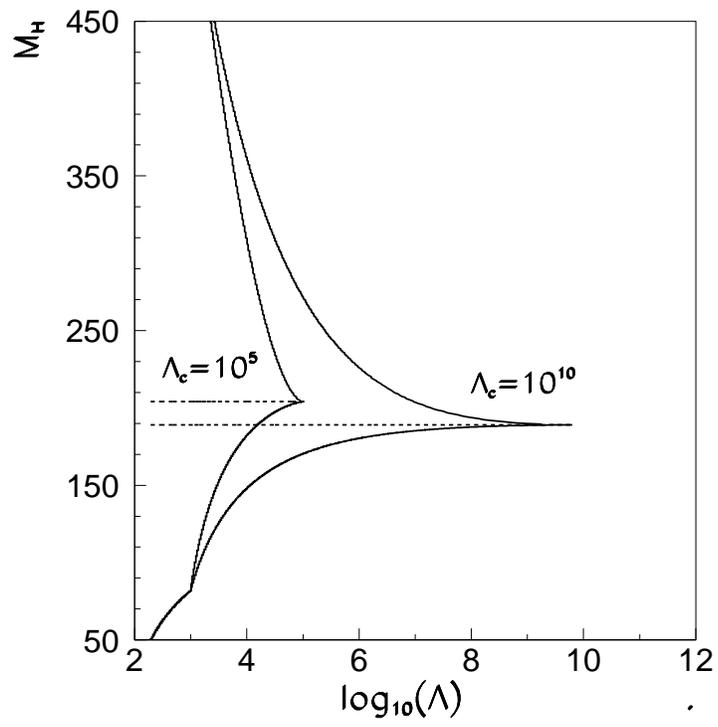}}
\vspace*{-10mm} 
\caption{ \label{fig3}%
 IR--UV fixed point structure and compositeness.
In the $\Lambda_c=10^5$~GeV case, $N_d=N_c=3$; 
In the $\Lambda_c=10^{10}$~GeV case, $N_d=N_c=1$. $M=10^3$~GeV.}
\end{center}
\end{figure}
\end{newpage}
\end{document}